\documentstyle[12pt,aasms4]{article}

\begin{document}

\noindent
{\it Dissertation Summary}

\begin{center}

%%% put title of your dissertation in following line:

\title{\large \bf A Search for Eclipsing Binaries in Galactic Globular
Clusters }

\end{center}

%%% Your name and current address below:

\author{ Kaspar von Braun }

\affil{ Department of Terrestrial Magnetism / Carnegie Institution of 
Washington }

\begingroup

\parindent=1cm

%%% supply the following information:

\begin{center}

Electronic mail: kaspar@dtm.ciw.edu

Thesis work conducted at: Dept of Astronomy, University of Michigan

Ph.D. Thesis directed by: Mario L. Mateo ;  ~Ph.D. Degree awarded: July 2002

%%% do not change following line:

%{\it Received \underline{\hskip 5cm}}

\end{center}

\endgroup

%%% fill in appropriate keywords from the list on the PASP web site
%% (http://pasp.phys.uvic.ca):


%%%place the text of your Dissertation Summary here:

Variable stars have historically served as tools and
laboratories in our understanding of stellar formation and
evolution, the formation of star clusters, the calibration of
distance determination methods, and a variety of other areas. In
particular, the study of eclipsing binaries (EBs) in globular
clusters (GCs) can provide direct distance estimates to clusters
as well as constraints for the turnoff masses of GC stars.
Knowing precise distances to GCs would constitute an independent
check of widely used distance determination methods in astronomy.
Obtaining masses of GC turnoff stars provides a fundamental test
of low-metallicity stellar models which calculate ages for these
stars - and thus the GCs themselves - and thereby allow a lower
limit estimate of the age of the universe.

My dissertation research with M. Mateo at the University of Michigan
consists of a monitoring survey of 10 Galactic GCs with the aim of
identifying photometrically variable EBs around or below the
main-sequence turnoff (MSTO). My thesis work comprises the results of
our research of the GCs NGC 3201, M10, and M12. Our observing
strategy is aimed at detecting variables with periods between 0.2 to
around 5 days and $16.5 < V < 20$, thereby optimizing our chance to
find the valuable detached EBs in the target clusters. In addition, we
correct for differential reddening variations across the cluster
fields of up to several tenths of magnitudes in $V$ by internally
creating a differential extinction map with a arcmin resolution from
our cluster photometry data, and calculate the additional reddening
zero point using isochrone fitting. The resulting improvement in the
appearance of the color-magnitude diagrams (CMDs) of the clusters is
considerable (see von Braun \& Mateo 2001, AJ, 121, 1522).

We obtained approximately 200 $VI$ epochs with about 20000 stars per
image for each of the three clusters. Analysis of these data revealed
the existence of 14 variable stars (11 EBs) in the field of NGC 3201,
three variables (1 EB) in the field of M10, and 2 EBs in the field of
M12. Spectroscopic follow-up work showed that only one variable (a
blue straggler W Ursa Majoris contact EB) in the field of NGC 3201 is
associated with the cluster. Another W UMa EB is most likely a member
of M12, based on its location in M12's CMD and its empirically
calculated absolute magnitude. The rest of the variable stars we
detected are members of the Galactic disk (see von Braun \& Mateo
2002, AJ, 123, 279; and von Braun et al. 2002, AJ, 124, 2067 for
details). We thus calculate a ratio of observable, short-period,
main-sequence binaries to main-sequence stars of around 1/500 for the
Galactic disk (consistent with literature estimates) and 1/9000 for GC
members. The latter value is considerably smaller than the commonly
quoted value of 1/1000.  While the discrepancy may be due to
small-number statistics, we attribute it to the fact that we lose
about 20\% of our fields to crowding toward the cluster center where
the binary fraction should be highest after a few cluster relaxation
times.

In my dissertation, I show the cluster fields and CMDs (before and
after dereddening) with the locations of the variable stars, our
differential extinction maps, as well as the phased lightcurves
and spectra (whereever applicable) of the variable stars. For
additional information, or to obtain a copy of the dissertation,
please contact me or visit http://www.ciw.edu/vonbraun.

I would like to express my sincere gratitude to the U of Michigan
Dept of Astronomy faculty (especially M. Mateo), postdocs (past
and present), and particularly my fellow graduate students
(especially K. Chiboucas) for their help, advice, and support.

%%% if there are references, place below, following standard reference style,
%% or to save space you may use "in-line" references within the text and delete
%% the lines below:
%
%\begin{references}
%
%\reference{  XXXX, Y.Z. 1998, ApJ, 234, 456}  
%
%\end{references}
%
\end{document}